# Current-limiting challenges for all-spin logic devices

Li Su, [1,2,3] Youguang Zhang,[1] Jacques-Olivier Klein[2,3], Yue Zhang,[1] Arnaud Bournel,[2,3] Albert Fert[1,4] and Weisheng Zhao[1,2,3,*]

[1] *Fert Beijing Institute, Univ. Beihang, 100191, Beijing, China*

[2] *Institut d'Electronique Fondamentale, Univ. Paris-Sud, F-91405 Orsay, France*

[3] *UMR 8622, CNRS, F-91405 Orsay, France*

[4] *Unité Mixte de Physique CNRS-Thales, F-91767 Palaiseau, France*

[*]Email: weisheng.zhao@buaa.edu.cn

### Abstract

All-spin logic device (ASLD) has attracted increasing interests as one of the most promising post-CMOS device candidates, thanks to its low power, non-volatility and logic-in-memory structure. Here we investigate the key current-limiting factors and develop a physics-based model of ASLD through nano-magnet switching, the spin transport properties and the breakdown characteristic of channel. First, ASLD with perpendicular magnetic anisotropy (PMA) nano-magnet is proposed to reduce the critical current ($I_{c0}$). Most important, the spin transport efficiency can be enhanced by analyzing the device structure, dimension, contact resistance as well as material parameters. Furthermore, breakdown current density ($J_{BR}$) of spin channel is studied for the upper current limitation. As a result, we can deduce current-limiting conditions and estimate energy dissipation. Based on the model, we demonstrate ASLD with different structures and channel materials (graphene and copper). Asymmetric structure is found to be the optimal option for current limitations. Copper channel outperforms graphene in term of energy but seriously suffers from breakdown current limit. By exploring the current limit and performance tradeoffs, the optimization of ASLD is also discussed. This benchmarking model of ASLD opens up new prospects for design and implementation of future spintronics applications.

Subject areas: Spintronic devices, Nanoscience and technology, Graphene.

### Introduction

According to the well-known Moore's law, the development of electronic device is undergoing the bottleneck of the power and performance with continuous minimization[1, 2]. Spintronics manipulates the electron spin instead of charge as state variable for electrical information processing, which gives rise to the possibility of many applications such as ultra-low power logic and non-volatile storage[3-6]. Spintronics devices have been proposed to perform logic operations, but most of them suffer from the large dynamic power inherent in the magnetic field to be used or in the requirement to frequently transform data between electrical and magnetic states for pipeline computing[7-9]. All-spin logic device (ASLD) is one of the most promising candidates to overcome the above issues since it both stores and computes with spin information and logic-in-memory structure[10]. The essential physical basis of ASLD is nonlocal spin transfer torque (STT) effect[11, 12] attributing to pure spin injection and detection in the lateral nonlocal spin valve (LNLSV)[13]. This phenomenon has been observed experimentally in all-metal LNLSV[14-17], generally copper is considered as typically channel material. Recently, a benchmarking methodology based on copper channel ASLD for computing system has been further introduced and simulated[18]. Furthermore, ASLD with perpendicular magnetic anisotropy (PMA) nano-magnet is suggested to lower STT switching critical current density[18, 19]. Nonetheless, with shrinking device dramatically, the current density challenges to exceeding the breakdown limit of copper due to electromigration (EM) crisis in 2015[20]. For nanometer size ASLD, the breakdown characteristic of metal channel cannot be ignored and new material is also in great demand.

With the features of high electronic mobility, weak spin orbit coupling and hyperfine interactions, graphene has attracted considerable interests for future spintronics material[21-23]. In addition, the breakdown current density of graphene is at least two orders magnitude larger than that of copper[24-27]. Moreover, nonlocal STT effect in graphene-based LNLSV has been experimental demonstrated[28, 29], subsequently graphene-based all-spin logic gate (G-ASLG) with PMA nano-magnet has been proposed and evaluated[19]. However, the current-limiting factors of ASLD are lacking in detailed analysis, especially the current conditions, which is very important to assess the feasibility and optimization.

In this paper, we investigate the current-limiting factors of ASLD and develop a physics-based model including nano-magnet switching, spin transport properties and breakdown characteristic of channel. First, we present different structures of ASLD. Second, we introduce the compact model of ASLD based on STT effect in PMA nano-magnet, spin transport properties in LNLSV, and breakdown current density of channel, such as graphene and copper. Finally, we address the current-limiting conditions and energy dissipation to assess and optimize ASLD, which contributes to the design and implement of future spintronics devices.

**Results and Discussions**

**All-spin logic device (ASLD)**. Firstly, as Fig.1 shows, main part of ASLD is the structure of LNLSV, which is composed of PMA input and output nano-magnet (F1 and F2) connected by a nonmagnetic channel (such as copper or graphene). It is based on spin transport properties in LNLSV and nonlocal STT switching to perform logic operation. Once a charge current, $I_{inj}$, is injected into the device, spin current beneath input nano-magnet can diffuse in both directions, i.e., toward leftside (as a spin and charge current) and toward rightside (as a pure spin current). Then a voltage $V_{det}$ can be measured in the output nano-magnet as a result of spin accumulation in parallel (P) or antiparallel (AP) alignments of magnetizations, and the spin signal is defined as $R_s=(V_{det}^{AP}-V_{det}^{P})/I_{inj}$. The spin current following into the output nano-magnet $I_{det}$ is used to switch its magnetization based on nonlocal STT effect. Thus ASLD stores information as the magnetizations of input and output and communicates through pure spin current, just as its name implies.

To alleviate the conductance mismatch problem[30-32] for graphene channel LNLSV, tunnel barrier has been added, seen in Fig.1 (a), to enhance the spin injection or detection efficiency $P_{1,2}$ and contribute to large spin signal $R_s$. Besides, asymmetric structure is presented as Fig.1 (b), where tunnel barrier is only added in the input. It can ensure high spin injection efficiency and large absorption with low contact resistance. For metal channel LNLSV or all-metal ASLD proposal, the transparent contact is generally used instead of tunnel barrier, as shown in Fig.1(c). In the following, the performance of current-limiting and energy with these three structures are compared and discussed.

**Current-limiting Models of ASLD**. Afterwards, in order to study the current-limiting factors of the ASLD, we have developed a physics-based compact model that integrates STT switching of output nano-magnet, spin transport properties of LNLSV, and breakdown characteristic of channel. It allows us to analyze parameters such as device dimension, material parameters, and contact resistance and their interdependences, as well as to explore how to overcome current-limiting challenges and optimize ASLD performance.

**Critical current of switching nano-magnet**. Firstly, compared to in-plane magnetic anisotropy, PMA nano-magnet can reduce critical current density $J_{c0}$ or current $I_{c0}$ due to absence of the easy-plane anisotropy term[11]. Therefore we consider ASLD with PMA nano-magnet so as to reduce required current essentially. In addition, the perpendicular anisotropy energy density $K_\perp$ of PMA is high enough to ensure high thermal stability $\Delta = E_b/k_B T$, where $E_b$, $k_B$, $T$ are energy barrier, Boltzmann constant and temperature, respectively. Regarding the nano-magnet size below the domain wall width, we assume single-domain magnetization reversal with STT effect[12, 13, 33]. Since size effect of demagnetization factors is considered based on macrospin model[33], $J_{c0}$ for PMA output nano-magnet can be derived as follows

$$J_{C0} = \frac{4e\alpha_{\text{eff}} E_b}{\hbar P A_c} = \frac{4e\alpha_{\text{eff}} K_{\text{eff}} t_F}{\hbar P} = \frac{2\alpha_{\text{eff}} e(2K_\perp - N_Z \mu_0 M_S^2) t_F}{\hbar P} \quad (1)$$

where $M_S$ is saturation magnetization and $P$ is spin polarization factor. $A_c$ is nano-magnet area and $t_F$ is free layer thickness of output nano-magnet. e, $\hbar$, $\mu_0$ are the elementary charge, reduced Planck constant and vacuum permeability, respectively. The demagnetization factors $N_z$ is calculated to decrease with shrinking the nano-magnet size[34], consequently it enhances the effective perpendicular energy density $K_{\text{eff}} = K_\perp - N_z \mu_0 M_s^2 / 2$ shown in Fig.2 (a). Satisfying that $J_{c0}$ is independent of nano-magnet area or only depends on nano-magnet thickness in the macrospin model,[33] we can obtain effective damping constant $\alpha_{\text{eff}}$ shown in Fig.2(a). Then the time-dependent magnetization dynamic is governed by the Landau-Lifshitz-Gilbert equation including spin torque[11, 12]. As a result, we give static and dynamic properties of switching output nano-magnet with scaling down (nano-magnet width $W$ from 30nm to 4nm) in Fig.2. It shows that thermal stability $\Delta$ scales almost linearly with nano-magnet width $W$, but critical current $I_{c0}$ decreases in proportion to nano-magnet area $A_c$, leading to increasing STT efficiency $\Delta/I_{c0}$ with shrinking nano-magnet. The above modified macrospin model is able to demonstrate STT behavior with size effect, whose theoretical results agree well with experimental observations[33, 35]. Note that, as shown in Fig.3(c), the critical current for required switching time 2ns, $I_{t_{sw}=2ns}$, which is equal to $I_{det,\ t_{sw}=2ns}$, is utilized to further estimate the performance of ASLD in this paper.

**Spin transport properties of ASLD.** Most importantly, spin transports properties are figured out to diminish input charge current $I_{inj}$ that generates $I_{det}$. Based on spin-dependent and one-dimensional drift-diffusion theory, we define and calculate spin transport efficiency $\eta$ as[13, 19]

$$\eta \equiv I_{det}/I_{inj} = 2e^{-L/\lambda_N}\left(\frac{P_1\frac{R_1}{R_N}}{1-P_1^2}+\frac{P\frac{R_F}{R_N}}{1-P^2}\right)\times\left[\prod_{i=1}^{2}\left(1+\frac{2\frac{R_i}{R_N}}{1-P_i^2}+\frac{2\frac{R_F}{R_N}}{1-P^2}\right)-e^{-2L/\lambda_N}\right]^{-1} \quad (2)$$

where $R_F = \rho_F\lambda_F/(t_F A_C)$ is nano-magnet spin resistance with spin diffusion length $\lambda_F$, resistivity $\rho_F$, thickness $t_F$ and contact or nano-magnet area $A_C$. $R_N = \rho_N\lambda_N/t_N W$ is spin resistance of nonmagnetic channel, where $\rho_N$, $W$, $t_N$ and $\lambda_N$ are the resistivity, width, thickness and spin diffusion length of channel, respectively. $R_i$ is contact resistance of the injector ($i$=1) or the detector ($i$=2). $P$ and $P_i$ are the spin polarization of the electrode and the interfaces, respectively. Eq.(2) and Fig.3 show that spin transport efficiency $\eta$ depends strongly on the material parameters ($R_i$, $P_i$, $\lambda_N$) as well as the device geometry (channel width $W$ and length $L$ shown in Figure.1). Especially, as shown in Fig.2 (a), higher efficiency can be obtained by increasing input contact resistance $R_1$ and decreasing output contact resistance $R_2$, thus asymmetric structure (as Fig.1 (b)) is expected. Note that, $R_iW$ type of contact resistivity is generally utilized for graphene instead of $R_iA$ type.[36] According to nano-magnet switching and spin transport properties of channel, the lower limit of $I_{inj}$ for workable ASLD is deduced as $I_{c0}/\eta$.

**Breakdown current density of ASLD.** Furthermore, with the object of finding the upper limit of $I_{inj}$ of ASLD, we study the breakdown current density of channel $J_{BR}$. In this work, we focus on typically spin channel materials, graphene and copper. For graphene channel due to Joule heating mechanism, we apply the size-dependent model of Liao et al.[37] Taking into account both heat loss to the substrate and to the interface, the breakdown current density of graphene channel $J_{BR,\,G}$ is calculated as [37]

$$J_{BR,G} = \left[ \frac{g(T_{BD}-T_0)}{\rho_G t_G W} \times \frac{\cosh(\frac{L}{2L_H}) + gL_H R_T \sinh(\frac{L}{2L_H})}{\cosh(\frac{L}{2L_H}) + gL_H R_T \sinh(\frac{L}{2L_H}) - 1} \right]^{1/2} \quad (3)$$

In our calculation, $T_{BD}$ is the breakdown temperature (~873K oxidation in air); $T_0$ is ambient temperature (295K); $L_H = \sqrt{k_G W t_G / g}$ is thermal healing length, $k_G$ is thermal conductivity of graphene, and $R_T$ is the thermal resistance at the metal contacts. The thermal contact resistance per unit length from the graphene channel to substrate is calculated as [37]

$$g^{-1} = \left\{ \frac{\pi k_{ox}}{\ln[6(t_{ox}/W+1)]} + \frac{k_{ox}}{t_{ox}} W \right\}^{-1} + \frac{R_{Cox}}{W} \quad (4)$$

Where $k_{ox}$ is the thermal conductivity of substrate; $t_{ox}$ is the thickness of substrate; $R_{Cox}$ is thermal resistance of the graphene-substrate interface. Taking graphene-SiO$_2$ as example based on above model and experimental results[27], we require lower resistivity, smaller dimension and thinner substrate for higher $J_{BR,\,G}$ as illustrated in Fig.4. Based on Eq. (3) and Eq. (4), higher thermal conductivity ($k_{ox}$) or lower thermal resistance of graphene-substrate interface ($R_{Cox}$) can achieve higher breakdown current density. It is a possible solution to utilizing diamond substrates (higher $k_{ox}$) or graphene-BN substrates (lower $R_{Cox}$ due to smoother interface).

Going forward metal channel, electromigration (EM) is one of the key current density limitations with shrinking size of device rapidly. We consider EM-failure mechanism for size-dependent breakdown current density of copper channel $J_{BR,\,Cu}$ with Blech model, and the threshold product is described as[38, 39]

$$(J_{BR,\,Cu} \times L)_{th} = \frac{\Omega \Delta \sigma}{Z^* e \rho_{Cu}} \quad (5)$$

where $\Omega$, $\Delta\sigma$, $Z^*$ are the atomic volume of copper, the normal stress difference between channel ends, and effective charge of copper. The threshold product can be obtained experimentally, for example, 1500A/cm for $Cu/SiO_2$, then the length-effect design can be realized. Furthermore, the increasing resistivity of copper $\rho_{Cu}$ with narrow-width effect is also under consideration[41], resulting in decreasing $J_{BR,\,Cu}$ based on Eq.(5). Therefore calculated $J_{BR,\,Cu}$ with size-effect are presented in Fig.4 (c) and Fig.4 (d). It shows that, $J_{BR,\,Cu}$ will suffer from the great growing resistivity and benefit from the short-length effect. In contrast, graphene can sustain current density around $10^{10}$A/cm$^2$, 2 orders of magnitude higher than that of copper.

**Current-limiting conditions and optimization.** Finally, the whole current-limiting factors have been investigated as above, namely the current-limiting conditions of ASLD can be written as following,

$$I_{c0}/\eta \leq I_{inj} \leq I_{BR} = J_{BR} \times A_c \quad (6)$$

Given a workable injection current $I_{inj}$, we can obtain its switching time $t_{sw}$ by solving LLG equation. Note that the process time of ASLD is determined by the STT switching, and the time of spin current propagation is ignored.[2, 40] To demonstrate the entire performance of ASLD, the device energy dissipation per bit can be estimated as

$$E = I_{inj}^2 \times R_{in} \times t_{sw} \quad (7)$$

where $R_{in}$ ($\sim R_1$) is input resistance related to the input charge current path. Based on Eq.(2), as contact resistance $R_1$ increases, spin transport efficiency $\eta$ is higher, which can reduce injection current $I_{inj}$ so as to help lower energy. However, the increase of $R_1$ leads to the augmentation of energy. Thus contact resistance $R_1$ has optimal value for the minimum energy of ASLD. As mentioned, the LLG solution value of $I_{det,\ t_{sw}=2ns}$ for nano-magnet $W$=10nm, $t_F$=2nm is taken into account. Fig.5 shows the injection current and energy as a function of contact resistance for both graphene and copper channel in different lengths $L$, the breakdown current is also illustrated as the upper limitation. In Fig.5 (a) and Fig.5 (c), the injection current $I_{inj}$ exhibits a minimum, since increasing $R_1$ will reduce $\eta$ and boost $I_{inj}$ after the saturation of spin signal $R_s$ with $R_1 \gg R_N$. The advantage of the breakdown current density of graphene makes a broad workable current range for ASLD, that is to say, we can enhance injection current $I_{inj}$ for fast speed $1/t_{sw}$. But it is true that the spin resistance of graphene $R_G$ is larger than that of copper, the optimal contact resistance of graphene channel corresponds to be larger. As shown in Fig.5 (b) and Fig.5 (d), the optimized energy of graphene channel ASLD is almost ten times larger than that of copper channel ASLD. In addition, it is found that the optimal value of contact resistance increases as the channel length.

At last, we have demonstrated and analyzed all the performances of ASLD in different materials (graphene and copper) and structures as shown in Fig.6. The dependence of each parameter is analyzed. Thanks to spin transport efficiency, asymmetric structure is the optimal structure for lower injection current $I_{inj}$ to overcome current-limiting issue. To reduce injection current and energy dissipation in the same structure, smaller device with larger aspect ratio $W/L$, higher spin injection efficiency $P_1$, and longer diffusion length $\lambda_N$ are expected. For the copper channel, the device geometry and material parameters are strictly satisfied due to the breakdown current. For example, if we take 10nm wide copper channel with both transparent contacts, F1/Cu/F2, breakdown will always occur unless the channel length $L$ is within 20nm as shown in Fig.6 (a). Even so, the required spin injection efficiency must reach up to 0.5 at least and the minimal spin diffusion length is 200nm. Compared with copper, graphene can sustain higher breakdown current so that it can easily obtain an achievable ASLD. However, the energy performance of graphene channel ASLD is restricted by its large contact resistance. It is a possible solution to finding new tunnel material that realizes lower contact resistance as well as high efficiency.

**Conclusions**

In summary, we have developed a physic-based model for ASLD including PMA nano-magnet switching, spin transport properties and breakdown characteristic of channel. Its current-limiting factors, i.e., critical current of nano-magnet, spin transport efficiency and breakdown current of channel, have been investigated. In order to estimate the feasibility and performance of ASLD, the current-limiting conditions and energy dissipation have been addressed. Moreover, it has been shown that the contact resistance can be optimized for minimum energy. Regardless of the spin channel materials, the asymmetric structure is the most effective for ASLD in terms of current limitation. Copper channel outperforms graphene in term of energy but seriously suffers from the breakdown current limit. By exploring the current limit and performance tradeoffs, we can expect larger aspect ratio *W/L*, higher spin injection efficiency *P₁* and longer diffusion length $\lambda_N$. Our work is significant in the design and implementation of reduced ASLD facing current-limiting challenges, which gives a promising prototype for future spintronics applications.

**Methods**

**Statistic and dynamic model of switching nano-magnet.** Spin transfer torque (STT) effect is simulated based on the model deriving from Slonczewski and Berger[11,12]. The critical current density ($J_{c0}$) and switching time ($t_{sw}$) for PMA nano-magnet vector ($\vec{m}$) have been calculated by following macrospin model based on Landau-Lifshitz-Gilbert (LLG) equation with STT term. The main parameters are described and listed in Table I. $J_{c0}$ is given as Eq.(1), and the LLG equation reads

$$\frac{(1+\alpha_{\text{eff}}^2)}{\gamma}\frac{\partial \vec{m}}{\partial t} = -\mu_0 \vec{m} \times \vec{H}_{\text{eff}} - \alpha_{\text{eff}}\mu_0 \vec{m} \times (\vec{m} \times \vec{H}_{\text{eff}}) - \gamma a_J \vec{m} \times (\vec{m} \times \vec{p}) + \alpha_{\text{eff}} a_J (\vec{m} \times \vec{p}) \quad (8)$$

where $a_J$ is the strength of spin transfer torque, $a_J = \hbar PJ/2eM_s t_F$, $J$ is the current density, and $\vec{p}$ is a unit vector representing the direction of the magnetization of the input nano-magnet. As the size of device shrinks, we consider the nano-magnet width $W$ smaller than the domain wall width (e.g.30nm[35]) and assume that single-domain magnetization reversal takes place. Providing that interfacial and bulk magnetic anisotropies are intact, the effective perpendicular anisotropy field is expressed as

$$\vec{H}_{\text{eff}} = (-N_x M_S, -N_y M_S, 2K_\perp/\mu_0 M_S - N_z M_S) \cdot \vec{m} \qquad (9)$$

where $N_x, N_y, N_z$ are demagnetization factors, which depend on the nano-magnet dimensions and satisfy that $N_x+N_y+N_z=1$ .[34] As a result, the effective perpendicular anisotropy energy density $K_{\text{eff}} = K_\perp - N_z \mu_0 M_s^2/2$ increases with decreasing $W$ owing to decrease of $N_z$ .[34] Supposing that critical current density only relies on the material parameters and thickness of nano-magnet, we can obtain the calculated effective damping constant that reduces with $W$ below 30nm.

In addition, the disturbance caused by thermal agitation during the magnetization switching is supposed to be negligible, but the initial position of the nano-magnet vector is thermally distributed at a finite temperature. Here we use root square average value calculated by Eq.(10) as an initial angel $\theta$

$$\langle \theta_0 \rangle = \sqrt{k_B T/(\mu_0 M_s H_{\text{eff}} V)} \qquad (10)$$

In this work, we simulated time-dependent magnetization dynamics of 2-nm-thick nano-magnet width of 4nm~30nm by solving LLG equation and obtained the different critical current density or current for required switching time, e.g. $I_{t_{sw}=2\text{ns}}$.

**Spin transport model of LNLSV** The lateral non-local spin valve (LNLSV) is the key element of the ASLD. Aimed at the spin transport efficiency $\eta \equiv I_{\text{det}}/I_{\text{inj}}$, we analyzed the interface effects and spin accumulation based on the spin transport model of Takahashi[13], taking into account the spin dependent and the one-dimensional drift-diffusion theory.

**Physical model of copper resistivity** Regarding surface scattering of the conduction electrons and scattering due to grain boundaries, the resistivity is calculated as follows [41]

$$\rho_{Cu} = \rho_{Cu,\,bulk} \left\{ \frac{1}{3} \bigg/ \left[ \frac{1}{3} - \frac{\chi}{2} + \chi^2 - \chi^3 \ln\left(1 + \frac{1}{\chi}\right) \right] \right. \\ \left. + \frac{3}{8} C(1-p) \frac{1+TR}{TR} \frac{\lambda_{m,Cu}}{W} \right\} \quad (11)$$

$$\text{with } \chi = \lambda_{m,Cu} Q / d(1-Q)$$

where $p$, $Q$, $\lambda_{m,Cu}$, $d$ is the specularity parameter, the reflectivity coefficient at grain boundaries, the mean free path and the average distance between grain boundaries, respectively. $C$ is a constant with value 1.2 for rectangular cross section channel. The parameters are set as $p=0.49$, $Q=0.27$, $\lambda_{m,Cu}=45\,\text{nm}$, $d=80\,\text{nm}$ and the others are listed in Table I.

**Acknowledgments**

This work is supported by China Scholarship Council (CSC), the Innovation Foundation of Beihang University (BUAA) for PhD Graduates, and French project ANR-MARS, ANR-DIPMEM, PEPS-NVCPU. The authors thank Qi AN and Zhaohao WANG for fruitful discussions.

**Author contributions**

W.S.Z. and A.F coordinated the proposal and supervised the project. L.S. developed the model and made the calculation. L.S., W.S.Z., Y.G. Z, J.O.K. and A.F. analyzed the data. L.S., W.S.Z., Y.Z., A.B. and A.F. edited the manuscript. All authors reviewed and commented the manuscript.

**Competing financial interests**

The authors declare no competing financial interests.

**Figure legends**

**Figure 1. Main part of all-spin logic device (ASLD), i.e. structure of lateral nonlocal spin valve (LNLSV).** (**a**) Structure with tunnel barrier contact in both the input and output F1/T/C/T/F2. (**b**) Structure with tunnel barrier contact in the input and transparent contact in the output F1/T/C/F2. (**c**) Structure with transparent contact both in the input and output F1/C/F2. $J_s$ and $J_c$ are spin current density and charge current density, respectively.

**Figure 2. Static and dynamic properties for switching the output nano-magnet.** The critical current density $J_{c0}$ is calculated as 4.68 MA/cm$^2$ based on Eq.(1), and all the used parameters are listed in the Table I. (**a**) Calculated effective perpendicular magnetic anisotropy energy density $K_{eff}$ and effective damping constant $\alpha_{eff}$ as a function of nano-magnet width $W$. (**b**) Thermal stability $\Delta$ with respect to nano-magnet width $W$. (**c**) Calculation of the critical current $I_{c0}$ and current for 2ns switching time $I_{t_{sw}=2ns}$ as a function of square nano-magnet area $A_C$. (**d**) The dependence of spin-torque switching efficiency $\Delta/I_{c0}$ on nano-magnet width $W$.

**Figure 3. Calculation of spin transport efficiency $I_{det}/I_{inj}$ of ASLD with graphene channel based on Eq.(2).** The default device geometry is $(W, L) =$ (30nm, 100nm), and the other parameters of graphene are listed in Table I. (**a**) Calculation of $I_{det}/I_{inj}$ as a function of contact resistivity $R_1W$ and $R_2W$. (**b**) Calculation of $I_{det}/I_{inj}$ as a function of spin injection/detection efficiency $P_1$ and $P_2$. (**c**) Calculation of $I_{det}/I_{inj}$ as a function of device geometry $(W, L)$. (**d**) Calculation of $I_{det}/I_{inj}$ as a function of diffusion length of graphene $\lambda_G$.

**Figure 4. Breakdown current density $J_{BR}$ of graphene and copper.** (**a**) The breakdown current density vs resistivity for 32nm×80nm $(W\times L)$ graphene channel on 90nm thickness SiO$_2$. Thermal conductivity of graphene and SiO$_2$ are $k_G$=100 W·m$^{-1}$·K$^{-1}$ and $k_{ox}$=1.4 W·m$^{-1}$·K$^{-1}$, respectively. Graphene-oxide interface thermal resistance $R_{cox}$=10$^{-8}$ m$^2$·K·W$^{-1}$. (**b**) Calculated dependence of $J_{BR}$ on oxide thickness $t_{ox}$ and graphene channel width $W$ based on Eq.(3) and Eq.(4). The inset shows that percentage of contribution to total thermal resistance ($g^{-1}$) from the graphene-oxide interface and substrate, as a function of graphene channel width at $t_{ox}$=90nm. Parameters used are same as above. (**c**) Calculated dependence of $J_{BR}$ as a function of channel width. $L$=100nm, and the other parameters are same as Table I. The inset shows the resistivity of copper will dramatically increase for sub-100nm wide based on Eq.(11), resulting in the decrease of $J_{BR, Cu}$ based on Eq.(5). (**d**) Calculated dependence of $J_{BR}$ as a function of channel length.

**Figure 5. Analysis and optimization of injection current and energy based on contact resistance.** (**a**) Calculated injection current $I_{inj}$ for switching time $t_{sw}$=2ns, $I_{inj, t_{sw}=2ns}$ and input resistance $R_{in}$ as a function of contact resistance $R_1$ at different

values of channel length for structure F1/T/G/F2. The dash line stands for the corresponding breakdown current density $J_{BR, G}$. (b) Calculated energy for switching time $t_{sw}$=2ns, $E_{t_{sw}=2ns}$ as a function of contact resistance $R_1$ at different values of channel length for structure F1/T/G/F2. (c) Calculated $I_{inj, t_{sw}=2ns}$ and $R_{in}$ as a function of contact resistance $R_1$ at different values of channel length for structure F1/T/Cu/F2. $J_{BR, Cu}$ is illustrated as dash line. (d) Calculated $E_{t_{sw}=2ns}$ as a function of contact resistance $R_1$ at different values of channel length for structure F1/T/Cu/F2.

**Figure 6. Calculated injection current $I_{inj}$ and energy for switching time $t_{sw}$=2ns in different materials and structures considering the breakdown current density $J_{BR}$.** The transparent contact resistivity of copper channel is used as $R_iA$=0, and the other default parameters are listed in Table I. (a) and (b) give the calculated dependence $I_{inj, t_{sw}=2ns}$ and $E_{t_{sw}=2ns}$ on channel length $L$. (c) and (d) give the calculated dependence $I_{inj, t_{sw}=2ns}$ and $E_{t_{sw}=2ns}$ on channel width $W$. (e) and (f) give the calculated dependence $I_{inj, t_{sw}=2ns}$ and $E_{t_{sw}=2ns}$ on spin injection efficiency $P_1$. (g) and (h) give the calculated dependence $I_{inj, t_{sw}=2ns}$ and $E_{t_{sw}=2ns}$ on spin diffusion length $\lambda_N$.

**Table I**. Main parameters for Calculation

| Parameter | Description | Default value |
|---|---|---|
| $M_s$ | Saturation magnetization | $9.5\times10^5$ A/m |
| $K_\perp$ | Perpendicular magnetic anisotropy energy density | $6.4\times10^5$ J/m$^3$ |
| $t_F$ | Free layer thickness of output nano-magnet | 2 nm |
| $W, L, t$ | Width, length and thickness of channel | 10nm, 20nm, $t_G$=0.335nm |
| AR, TR | Aspect Ratio $L/W$, $t/W$ | 2, 6/5(copper) |
| $A_C$ | Nano-magnet area and contact area | $10\times10$ nm$^2$ |
| $\alpha_{eff}$ | Effective damping constant | 0.007 |
| $N_x, N_y, N_z$ | Demagnetization factors | 0.17, 0.17, 0.66 [a] |
| $P$ | Spin polarization factor | 0.5 |
| $P_i$ | Spin injection ($i$=1) or detection ($i$=2) efficiency | 50 %( 0~100%) |
| $\lambda_G, \lambda_{Cu}, \lambda_F$ | Spin diffusion length of graphene, copper and ferro nano-magnet | 3 μm, 400 nm, 5nm |
| $\rho_{Cu,bulk}, \rho_G, \rho_F$ | Resistivity of bulk copper, graphene channel and nano-magnet | 1.7 μΩ·cm, 0.3kΩ [b], 20 μΩ·cm |
| $R_i \cdot W$ | Contact resistivity for graphene channel in the input ($i$=1) or the output ($i$=2) | 10 Ω·μm (Tunnel), 1 Ω·μm(Transparent) |
| $R_i \cdot A$ | Tunnel contact resistivity for copper channel | 0.02 Ω·μm$^2$ |
| Constant | Description | Value |
| $\hbar$ | Reduced Planck constant | $1.054\times10^{34}$ J·s |
| e | Elementary charge | $1.6\times10^{-19}$ C |
| $\mu_0$ | Vacuum permeability | $1.2566\times10^{-6}$ H/m |
| $\gamma$ | Gyromagnetic ratio | $1.76\times10^{11}$ rad·s$^{-1}$·T$^{-1}$ |

[a] Calculated from $W$ and $t_F$ [34], [b] Normalized by $t_G$.

**(a)**

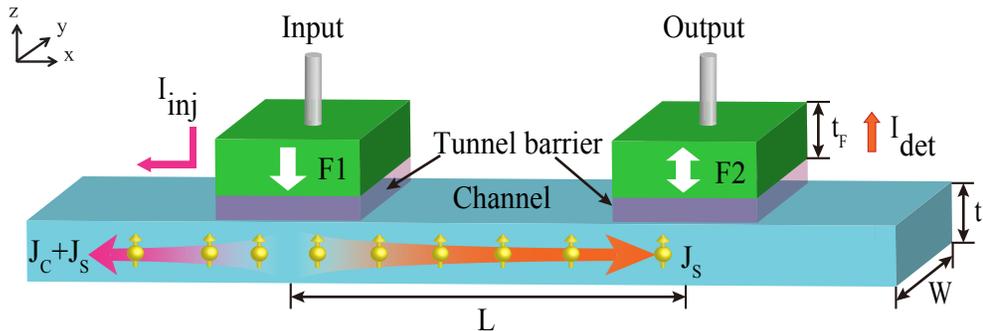

**(b)**

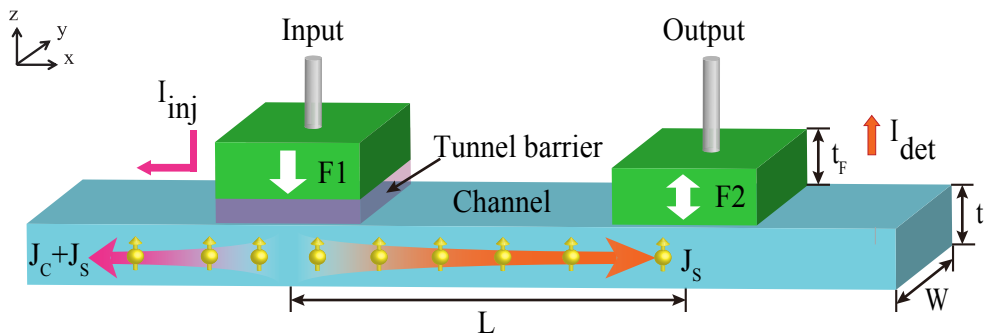

**(c)**

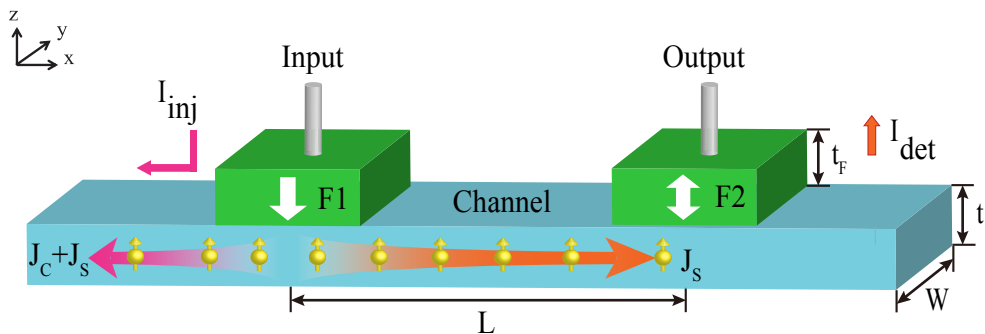

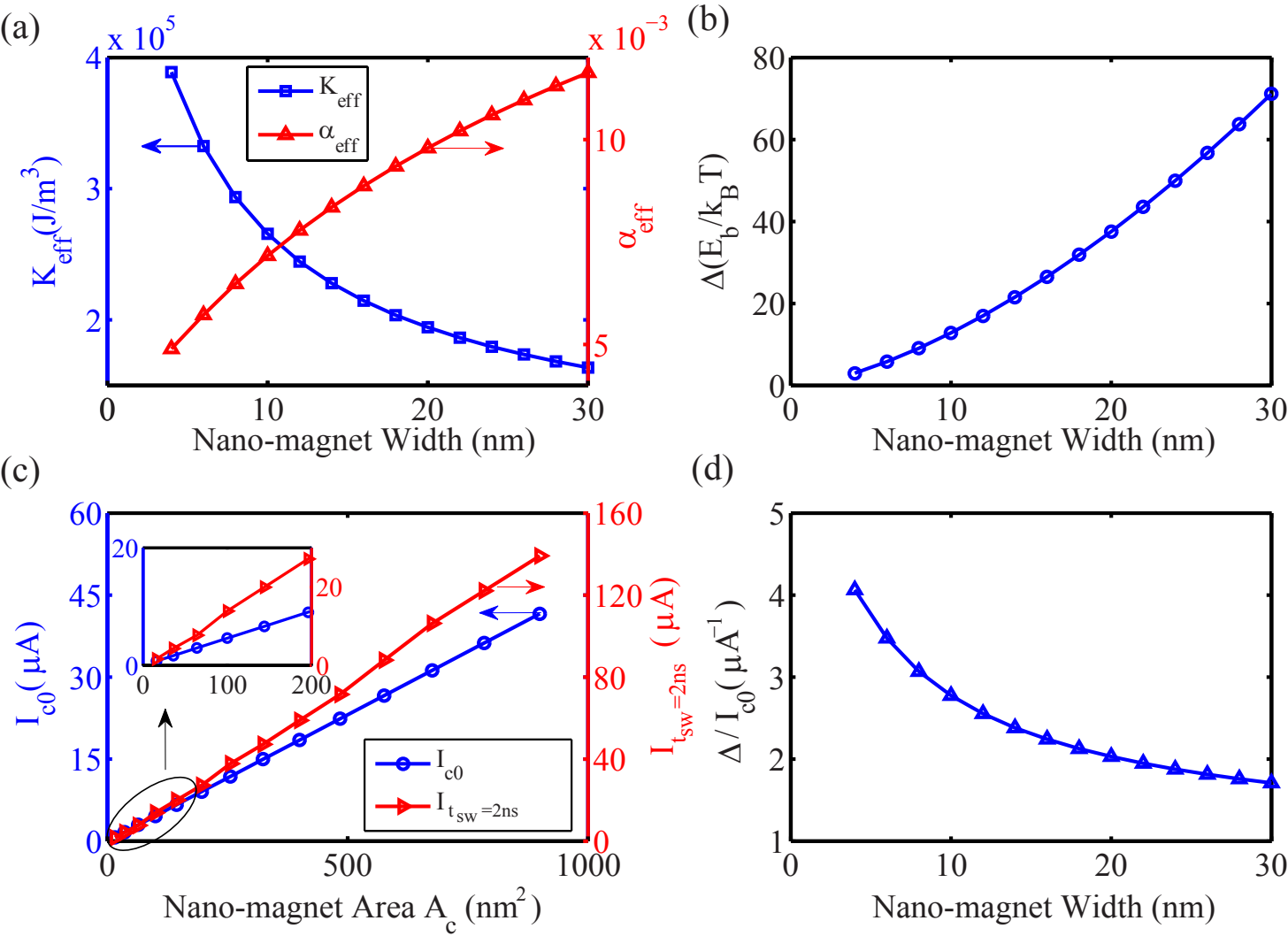

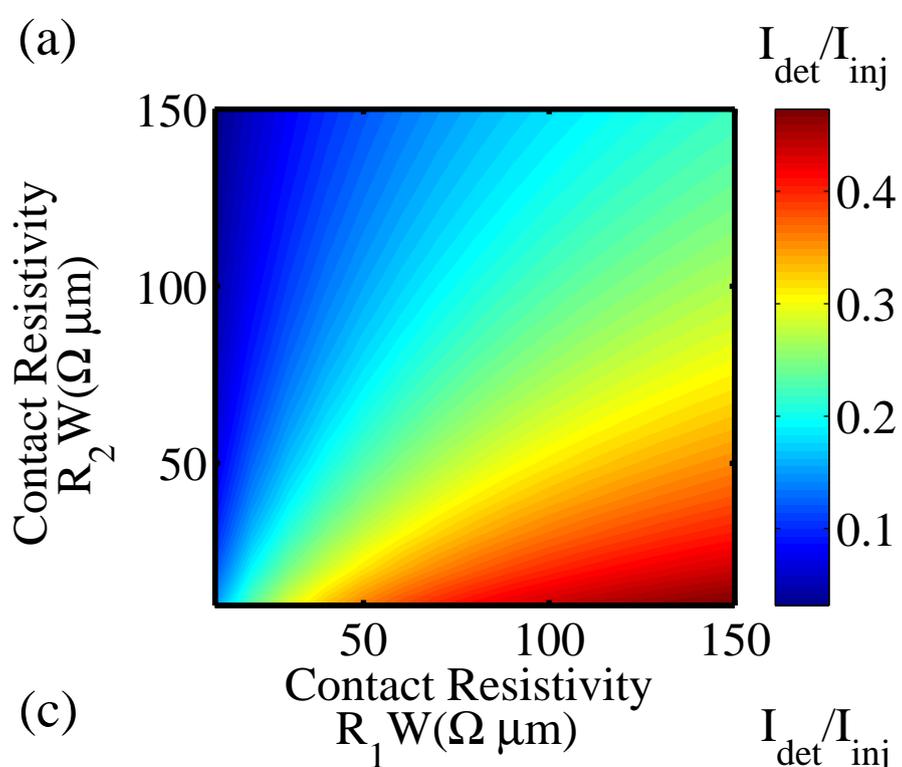 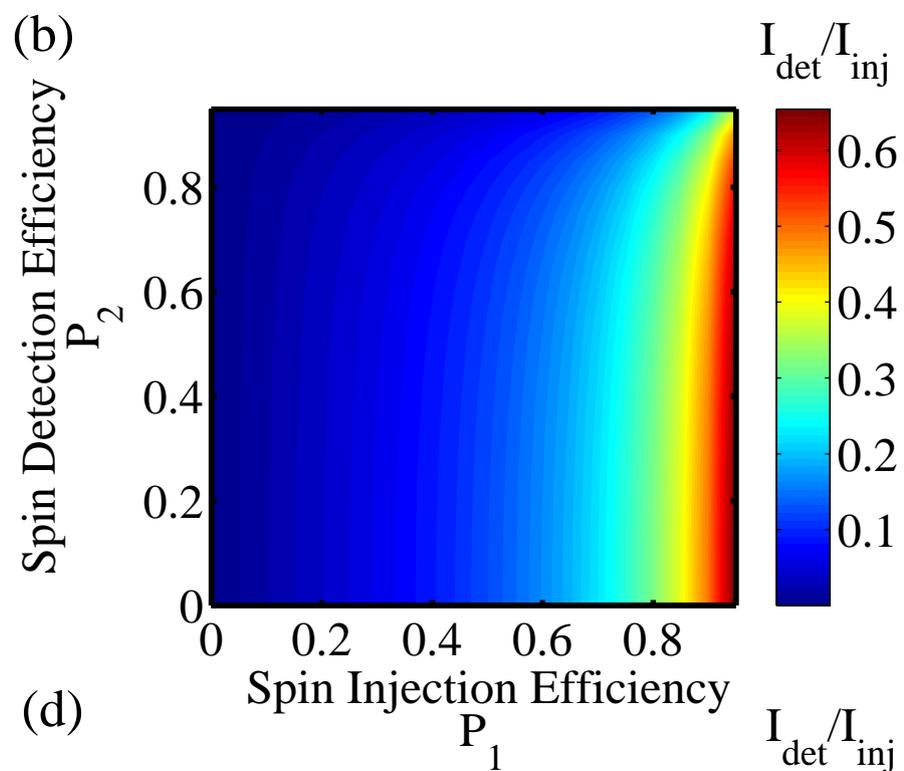 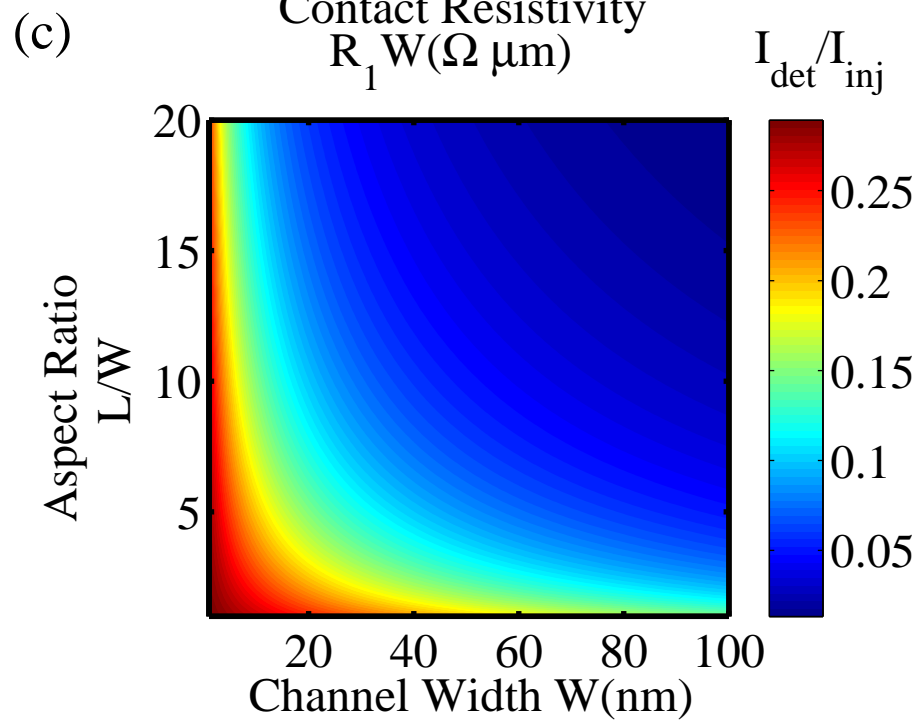 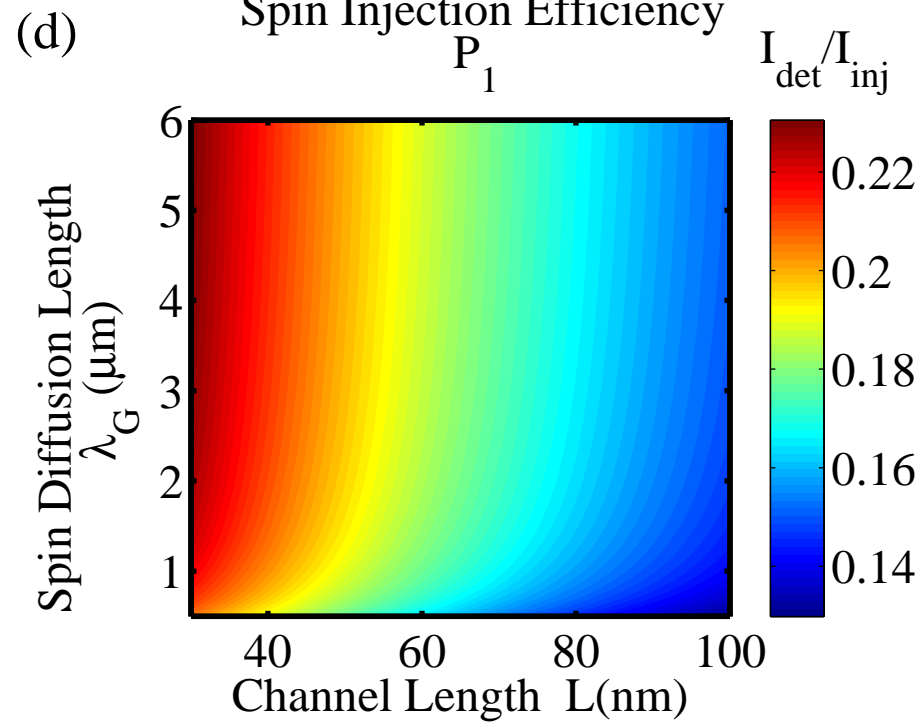

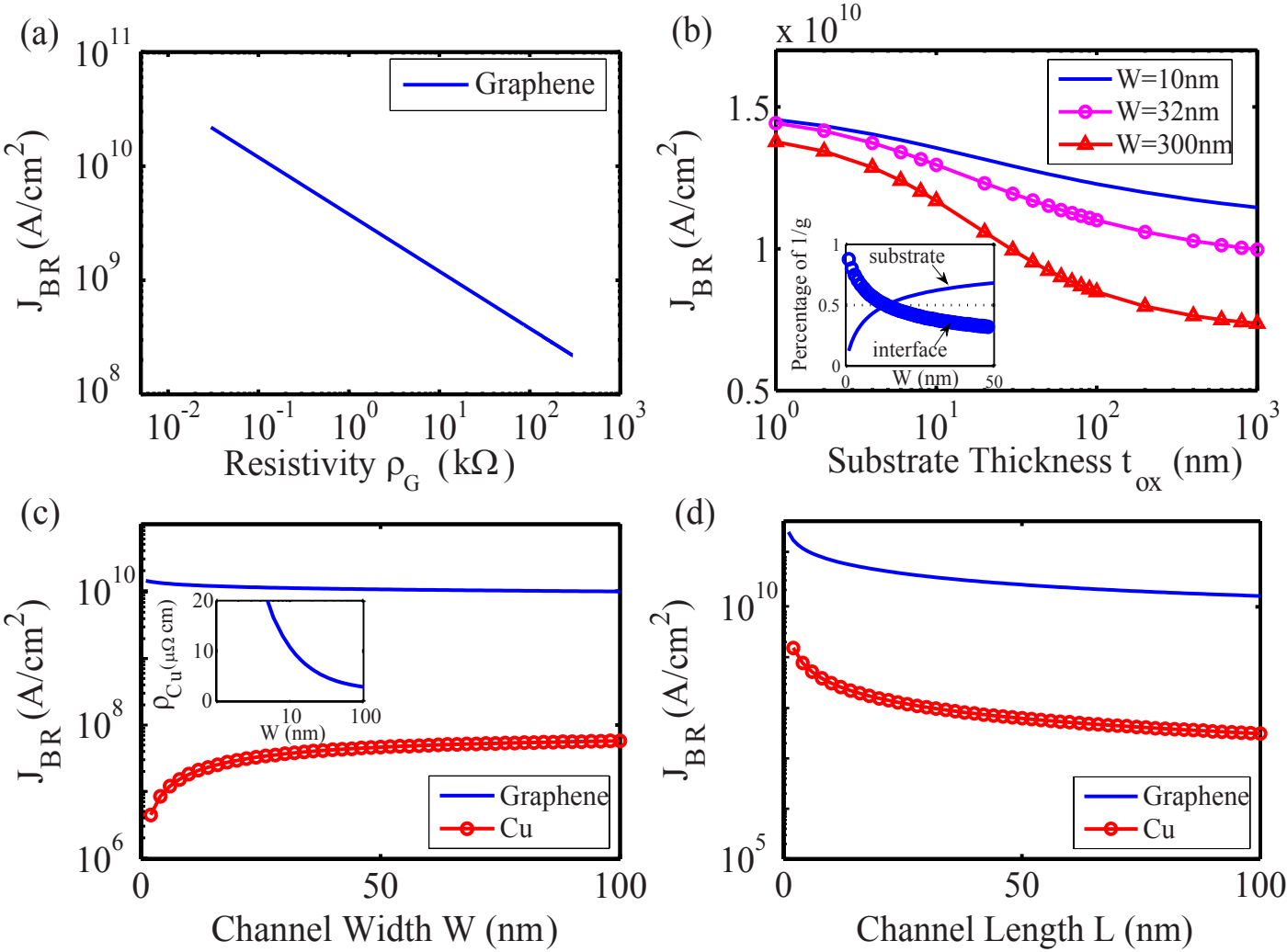

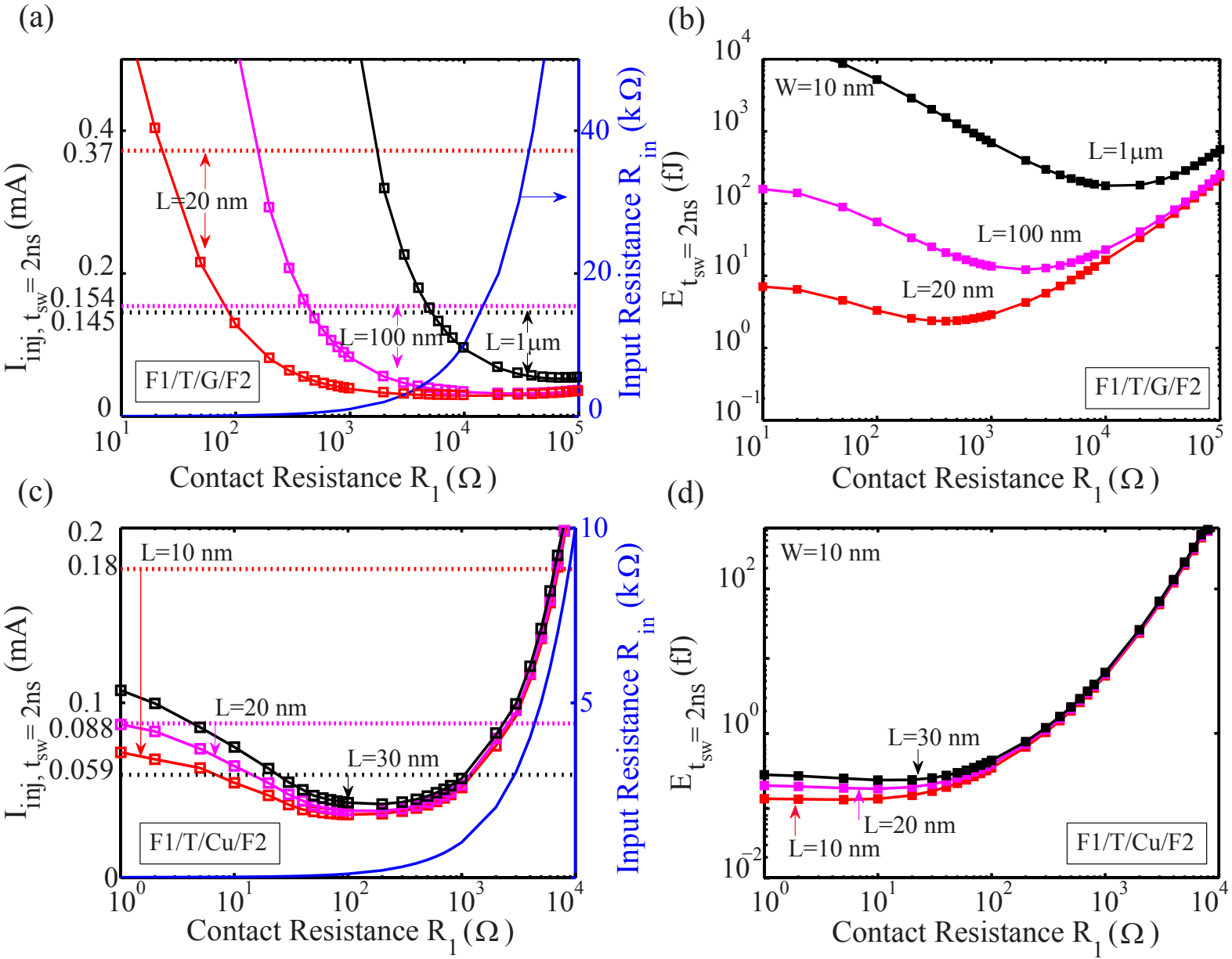

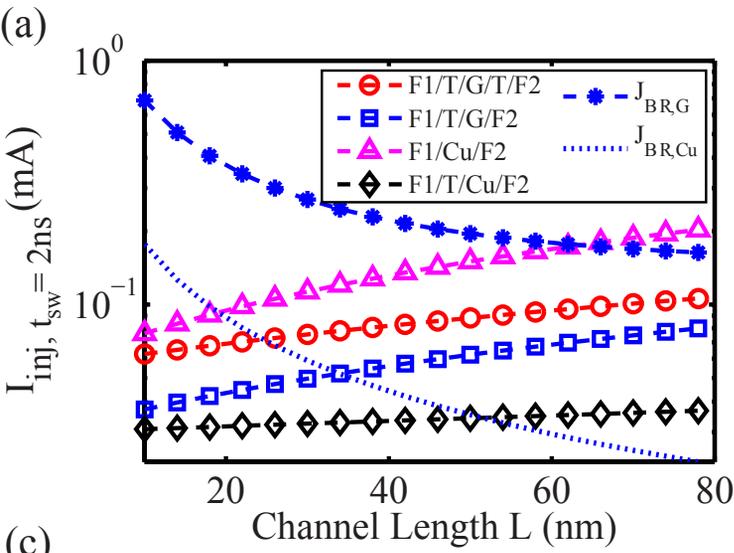
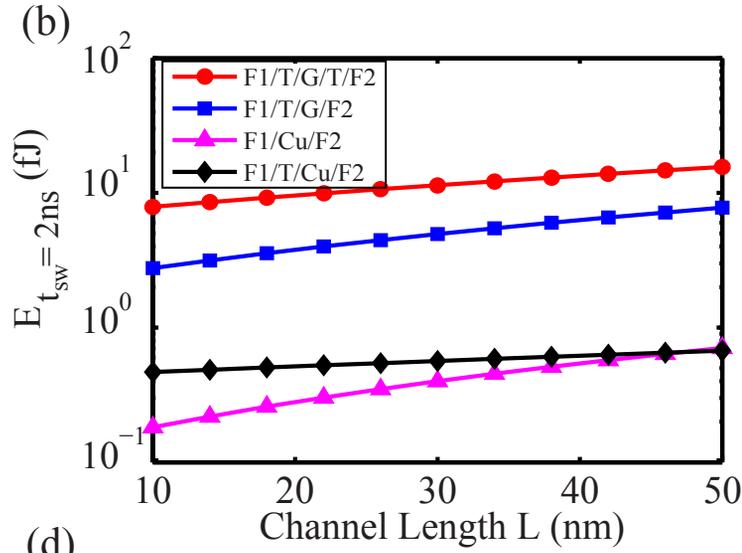
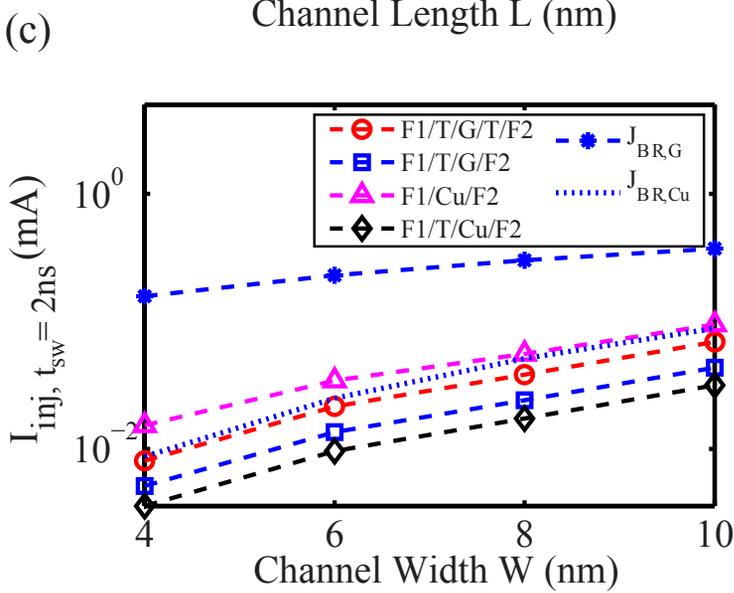
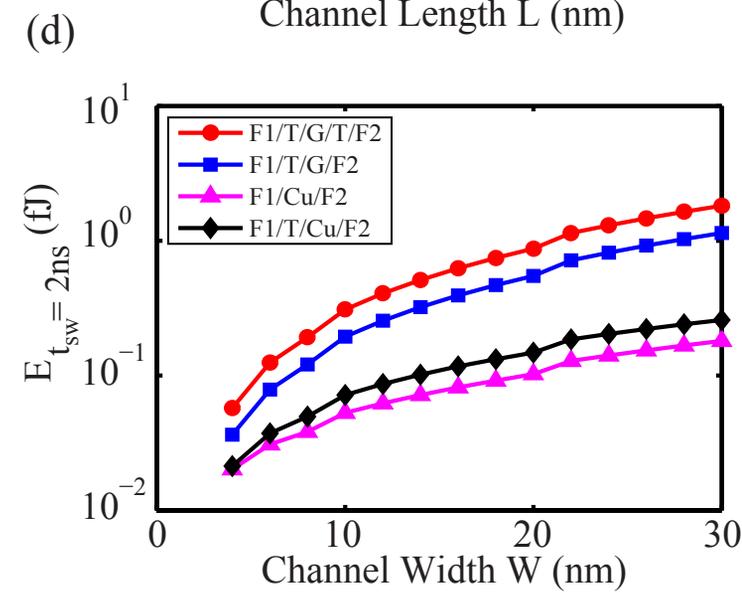
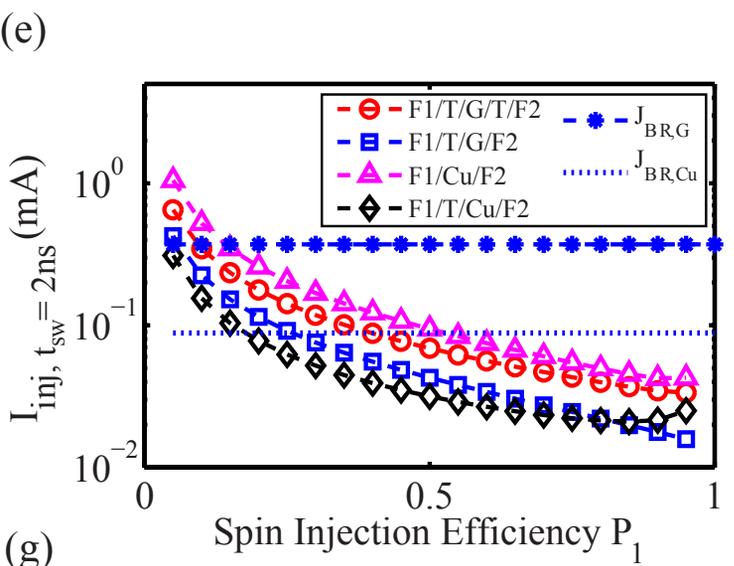
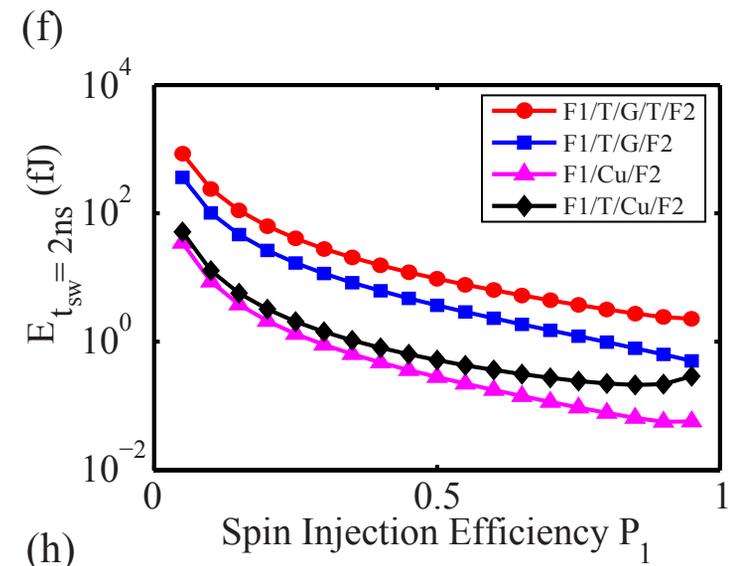
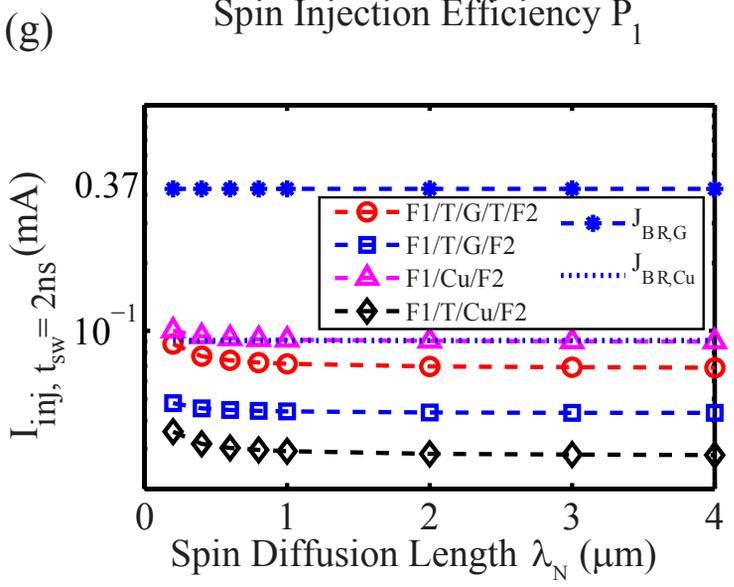
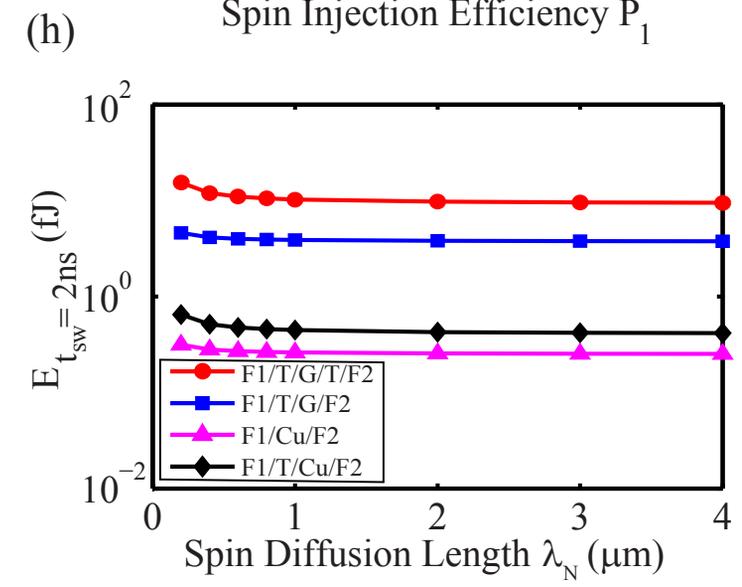